\documentclass[prb,twocolumn,amsmath,amssymb,superscriptaddress,longbibliography]{revtex4-2}

\usepackage[colorlinks,bookmarks=true,citecolor=blue,linkcolor=red,urlcolor=blue,citecolor=blue]{hyperref}
\usepackage{graphicx}
\usepackage{subfigure}
\usepackage{xcolor}

\begin{document}

\title{Luttinger Liquid phase in the Aubry-Andr\'{e} Hubbard chain}

\author{Runze Chi}
\affiliation{Beijing National Laboratory for Condensed Matter Physics and Institute of Physics, Chinese Academy of Sciences, Beijing 100190, China.}
\affiliation{School of Physical Sciences, University of Chinese Academy of Sciences, Beijing 100049, China.}

\author{Josephine J. Yu}
\affiliation{Department of Applied Physics, Stanford University, Stanford, California 94305, USA}

\author{Chaitanya Murthy}
\email{crmurthy@stanford.edu}
\affiliation{Department of Physics, Stanford University, Stanford, California 94305, USA}

\author{T. Xiang}
\affiliation{Beijing National Laboratory for Condensed Matter Physics and Institute of Physics, Chinese Academy of Sciences, Beijing 100190, China.}
\affiliation{School of Physical Sciences, University of Chinese Academy of Sciences, Beijing 100049, China.}
\affiliation{Beijing Academy of Quantum Information Sciences, Beijing, 100190, China.}

\begin{abstract}

We study the interplay between an on-site Hubbard repulsion and quasiperiodic potential in one-dimensional fermion chains using the density matrix renormalization group. We find that, at half-filling, the quasiperiodic potential can destroy the Mott gap, leading to a metallic Luttinger liquid phase between the gapped Mott insulator at strong repulsion and localized gapless Aubry-Andr\'{e} insulator at strong quasiperiodic potential. Away from half-filing, the metallic phase of the interacting model persists to larger critical strengths of the potential than in the non-interacting case, suggesting interaction-stabilized delocalization at finite doping. We characterize the Luttinger liquid through its charge and spin correlations, structure factors, and entanglement entropy.

\end{abstract}

\maketitle


Coulomb repulsion and disorder both serve as key driving forces for the metal--insulator transition (MIT) \cite{Mott1949,Mott1968,Hubbard1963,Anderson1958}. Understanding the interplay between these factors is a longstanding fundamental question in condensed matter physics \cite{Abrahams2001,Kravchenko2003,Belitz1994,Lee1985}. In the absence of disorder, Coulomb repulsion alone can lead to the formation of a Mott insulator. Conversely, in the absence of any interaction, disordered one-dimensional (1D) or two-dimensional (2D) systems always exhibit Anderson localization \cite{Abrahams1979}. When both interaction and disorder are present, one might expect that the tendency towards insulation is even stronger; however, the competition between these two factors may actually stabilize a novel metallic phase.

The possibility of such a metallic phase in 2D interacting disordered systems was first suggested by Finkelstein on the basis of renormalization group calculations \cite{Finkelstein1983,Finkelstein1984}. Subsequent transport experiments on 2D electron gases in silicon metal-oxide-semiconductor field-effect transistors provided compelling evidence for its existence \cite{Kravchenko1994,Kravchenko2003}. Furthermore, investigations employing the 2D Anderson-Hubbard model \cite{Denteneer1999,Heidarian2004,Karmakar2022,Lahoud2014,Szabo2020,Byczuk2005,Aguiar2009,Tanaskovi2003,Lee2016,Su2020} also provided support for the existence of this phase. Unlike a conventional Fermi liquid, the metallic phase was suggested to be an inhomogeneous percolating metal with insulating antiferromagnetic puddles \cite{Denteneer1999,Heidarian2004,Karmakar2022}.

In contrast, the 1D system exhibits a stronger tendency towards insulation. Even with infinitesimal disorder, the repulsive interacting system undergoes a transition into an insulating phase \cite{Giamarchi1988,Shankar1990}. Numerical findings indicate that the ground state of the disordered Hubbard chain is a charge-insulating random-singlet phase characterized by exponential decay of charge correlations but power-law ($1/r^2$) decay of spin correlations \cite{Yu2022}. No evidence of a metallic phase has been observed in the 1D model with uncorrelated disorder, with or without (repulsive) interactions. 

However, there exists a special type of deterministic disorder---a quasiperiodic potential---for which the non-interacting 1D system does host a metallic phase. In particular, the paradigmatic Aubry-Andr\'{e} (AA) model \cite{Aubry1980} exhibits a MIT at a critical strength $\lambda = \lambda_\mathrm{c}$ of the quasiperiodic potential; when $\lambda < \lambda_\mathrm{c}$ all single-particle states are extended, and when $\lambda > \lambda_\mathrm{c}$ all states are localized. In spinless fermion AA chains, it was observed that adding nearest-neighbor repulsive interactions causes $\lambda_\mathrm{c}$ to increase, resulting in a larger metallic region than in the non-interacting model \cite{Piero2016,Miguel2023}. This is reminiscent of the interaction-induced delocalization observed in the 2D Anderson-Hubbard model. 

Here, we investigate the ground state phase diagram of the spinful Aubry-Andr\'{e} chain with repulsive on-site Hubbard interactions, using the density matrix renormalization group method. We find that both the half-filled and hole-doped systems support an extended metallic phase, which appears to be stabilized by the competition between the quasiperiodic potential and repulsive interactions. We characterize the metallic state as a Luttinger liquid with $SU(2)$ spin symmetry by studying charge and spin correlations, as well as the entanglement entropy, but find signatures of the quasiperiodic potential in the spin structure factor and the electron distribution function. Our results may provide insights into interaction-induced delocalization in the 2D or 3D disordered problem, and suggest that the 1D Aubry-Andr\'{e}-Hubbard chain may capture other aspects of the interplay between disorder and correlations in higher dimensions.

\section*{Model and Method}

We study the 1D Aubry-Andr\'{e} Hubbard (AAH) model. The Hamiltonian is
\begin{equation}
\label{eq:Ham}
H = -t \sum_{i\sigma} (c_{i \sigma}^{\dagger} c_{i+1 \sigma}^{\phantom{\dagger}} 
+ \text{h.c.})
+ \sum_i V_i \, n_i
+ U \sum_i n_{i \uparrow} n_{i \downarrow} .
\end{equation}
Here $c_{i \sigma}^{\dagger}$ ($c_{i \sigma}^{\phantom{\dagger}}$) are the creation (annihilation) operators of spin-half fermions on a chain of $L$ sites, $n_{i\sigma} \equiv c_{i \sigma}^{\dagger} c_{i \sigma}^{\phantom{\dagger}}$ and $n_i \equiv \sum_{\sigma} n_{i\sigma}$ are the associated number operators, and
\begin{equation}
\label{eq:V_i}
V_i = \lambda \cos(q \, i + \phi)
\end{equation}
is a quasiperiodic potential with wavevector $q=2\pi b$, where $b$ is an irrational number (here we choose $b=\frac{2}{\sqrt{5}-1}$) and $\phi \in [0,2\pi)$ is a phase. We study the model at fixed electron density $n = \sum_i n_i /L$; half filling is $n=1$.

When $U=0$ in~(\ref{eq:Ham}), we recover the Aubry-Andr\'{e} model. It exhibits an $n$-independent localization transition at $\lambda/t=2$: all states are localized for $\lambda/t > 2$ or extended for $\lambda/t < 2$ \cite{Aubry1980}. When $\lambda=0$, we recover the 1D Hubbard model, which is exactly solvable by Bethe ansatz \cite{Lieb1968}. It is a Mott insulator at half filling for any $U>0$, and a metal at $n \neq 1$. When both $\lambda \neq 0$ and $U \neq 0$, the model cannot be solved exactly. Instead, we study it numerically.

We investigate the ground state phase diagram of the AAH model (\ref{eq:Ham}) with open boundary conditions using the density matrix renormalization group (DMRG) method \cite{White1992}, as implemented in the ITensor library \cite{ITensor,ITensor-r0.3}. We study the model both at half filling ($n=1$) and at a finite hole doping ($n=11/12$). Due to the presence of the quasiperiodic potential, the convergence of DMRG is relatively slow when compared to clean systems. To address this problem, we employ a combination of two-site and single-site DMRG techniques. Specifically, we perform 2 two-site steps followed by 6-10 single-site steps, and gradually increase the bond dimension in two-site steps to enhance convergence. To ensure accuracy in our results, we keep the truncation error $\leq 10^{-8}$ for all parameter values considered. Observables are either computed at a fixed phase $\phi=\pi/5$ of the quasiperiodic potential (\ref{eq:V_i}) or averaged over $10$ equally spaced phases, $\phi = 0, \pi/5, 2\pi/5, \dots, 9\pi/5$; we indicate all phase-averaged quantities by overbars.

\section*{Results}

\begin{figure*}[t!]
    \centering
    \includegraphics[width=\textwidth]{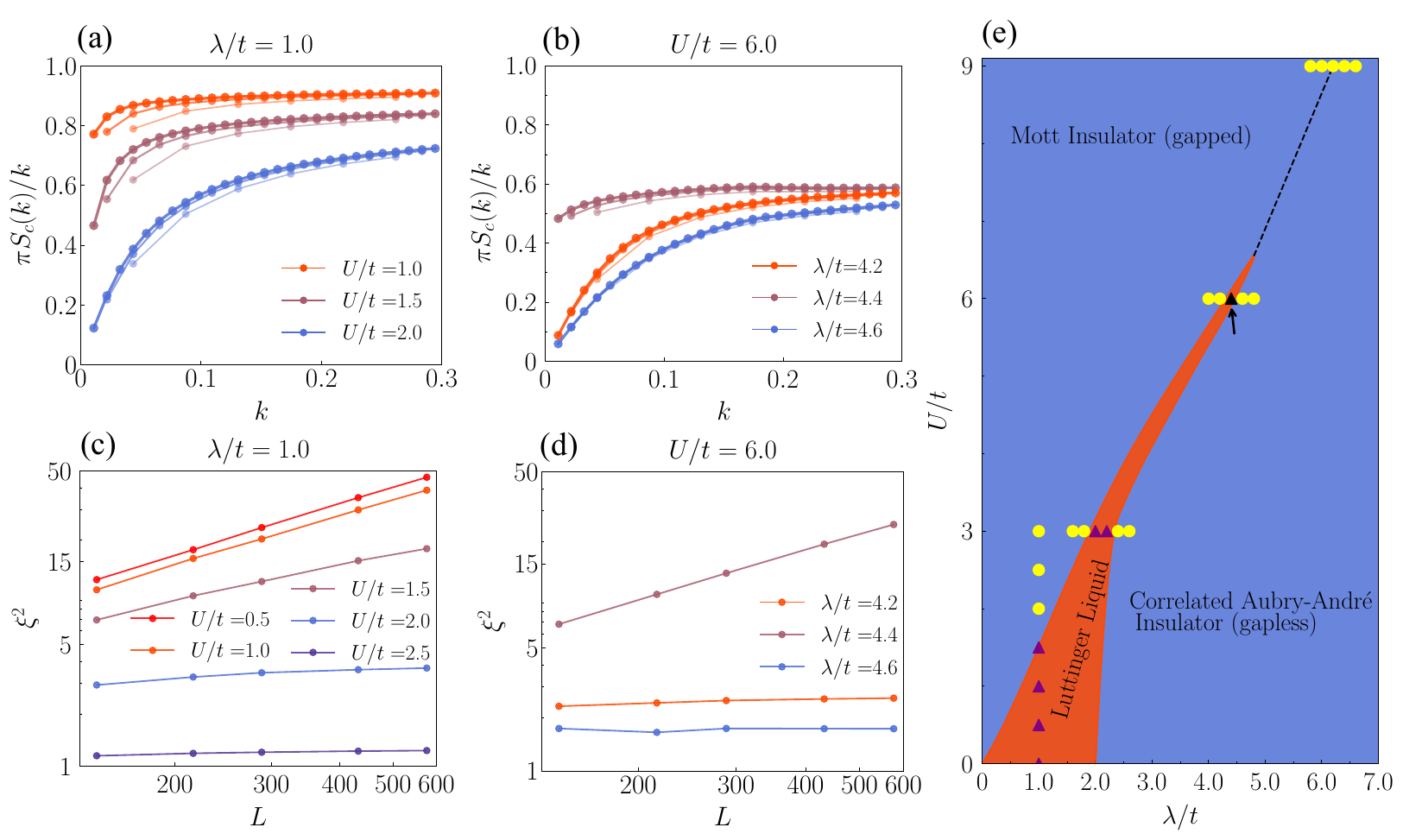}
    \caption{Results on the ground state phase diagram of the AAH model at half filling. 
    (a)--(b) Behavior of the charge structure factor $S_{\mathrm{c}}(k)$ as $k \to 0$. Heavier lines correspond to increasing system sizes of $L = 144$, $288$, and $576$ sites. A nonzero extrapolated intercept of $\pi S_{\mathrm{c}}(k)/k$ indicates a metal, while a vanishing intercept indicates an insulator, as discussed in the text.
    (c)--(d) Scaling of the charge localization length $\xi$ with system size $L$. $\xi^2 \sim L^{\alpha}$ with $\alpha > 0$ in the metal, while $\xi^2 \sim L^0$ in the insulator.
    Data are shown for various values of $U/t$ at $\lambda/t = 1$ (panels a, c) and at $U/t = 6$ for various values of $\lambda/t$ (panels b, d).
    (e) Estimated ground state phase diagram of the half-filled AAH model. Symbols mark parameter values at which data were collected. These are identified as metallic (purple/black triangles) or insulating (yellow circles) according to the behavior of $S_{\mathrm{c}}(k)$ and $\xi$ (both metrics yield consistent results). The Mott insulating phase is further distinguished by a nonzero charge gap $\Delta_{\mathrm{c}}$. The precise shape of the phase boundaries is conjectural, but is the simplest configuration consistent with our numerical results at the marked points and the known results in the $U=0$ or $\lambda=0$ limits.
    \label{fig1}}
\end{figure*}

We begin by investigating the phase diagram at half filling, $n=1$. The results are summarized in Fig.~\ref{fig1}. In the exactly solvable limits ($\lambda = 0$ or $U=0$), there are three well-understood phases: a Mott insulator at $\lambda = 0$ and $U > 0$ \cite{Lieb1968}, an Aubry-Andr\'{e} insulator at $\lambda > 2t$ and $U = 0$, and a metal at $\lambda < 2t$ and $U = 0$ \cite{Aubry1980}. We consider the effects of nonzero $\lambda$ and $U > 0$ on these phases and the boundaries between them.

We identify the metallic and insulating phases using two related metrics. The first is the long-wavelength ($k \to 0$) behavior of the charge structure factor
\begin{equation}
\label{eq:S_charge_k}
S_{\mathrm{c}}(k) = 
\frac{1}{L}\sum_{jl} e^{ik(x_j-x_l)}
\left[ \langle n_j n_l \rangle - \langle n_j \rangle \langle n_l \rangle \right] .
\end{equation}
Here $\langle \cdot \rangle$ denotes the ground state expectation value. Note that $S_{\mathrm{c}}(k) = S_{\mathrm{c}}(-k)$ because $\langle n_j n_l \rangle  = \langle n_l n_j \rangle$, and that $S_{\mathrm{c}}(0) = (\langle n^2 \rangle - \langle n \rangle^2)/L \equiv 0$ because we work at fixed electron density. In an insulator, we expect $S_{\mathrm{c}}(k) \propto k^2$ as $k \to 0$. This is because charge correlations in an insulator are short-ranged in real space, so the sum in (\ref{eq:S_charge_k}) converges absolutely and $S_{\mathrm{c}}(k)$ is an analytic function of $k$. By contrast, in a 1D metal, there are long-range charge correlations and we expect instead $S_{\mathrm{c}}(k) \sim (K_{\mathrm{c}}/\pi) \lvert k \rvert$ as $k \to 0$, where $K_{\mathrm{c}}$ is the charge Luttinger parameter.

We distinguish between these behaviors by plotting $\pi S_{\mathrm{c}}(k)/k$ versus $k$ for $k > 0$, and asking whether the intercept of the curve is zero (insulator), or a nonzero constant, $K_{\mathrm{c}}$ (metal). Representative plots are shown in Figs.~\ref{fig1}(a) and (b). Note that, in the metallic regime, there is actually a system-size--dependent scale $k^*(L)$ below which the curves bend downward due to finite-size effects. The systematic decrease of this crossover scale, $k^*(L)$, with increasing $L$ serves as a finite-size diagnostic of the metallic phase.

A second, closely related, metric for distinguishing the insulating and metallic phases is the ``charge localization length'' $\xi$,
\begin{equation}
\label{eq:xi_sq}
\xi^2 = - \frac{1}{2L} \sum_{jl} (x_j - x_l)^2 
\left[ \langle n_j n_l \rangle - \langle n_j \rangle \langle n_l \rangle \right] .
\end{equation}
Note that $\xi^2$ is formally the second derivative of $S_{\mathrm{c}}(k)/2$ at $k=0$. It diverges as a power of the system size in the metallic phase, $\xi^2 \sim L^{\alpha}$, and saturates to a finite constant in the insulating phase, $\xi^2 \sim L^0$. Representative plots of $\xi^2$ versus $L$ for different values of $\lambda/t$ and $U/t$ are shown in Figs.~\ref{fig1}(c) and (d). Benchmark results for both $S_{\mathrm{c}}(k)$ and $\xi^2$ in the non-interacting case are presented in Appendix~\ref{app:AA_benchmark} for comparison. 

Our results for the ground state phase diagram of the half-filled AAH model are summarized in Fig.~\ref{fig1}(e). The system is a Mott insulator at large $U$, and a correlated Aubry-Andr\'{e} insulator at large $\lambda$. These insulating phases are distinguished by, e.g., the charge gap,
\begin{equation}
\label{eq:Delta_c}
\Delta_{\mathrm{c}} = \tfrac{1}{2} E_{N+2} + \tfrac{1}{2} E_{N-2} - E_{N} ,
\end{equation}
which is nonzero in the Mott insulator but zero in the AA insulator (here $E_N$ is the ground state energy with $N$ electrons). At sufficiently large $\lambda/t$ and $U/t$, there appears to be a direct transition between the two insulators, which we estimate as the dotted line in Fig.~\ref{fig1}(e) (see Appendix~\ref{app:ins_transition} for details). At smaller $\lambda/t$ and $U/t$, we find a metallic phase sandwiched between the insulating phases. Notably, the metallic region extends along a narrow sliver in parameter space to larger values of $\lambda$ and $U$, suggesting that it is indeed stabilized by competition between opposing tendencies towards potential-driven localization and interaction-driven Mott insulation. The shape of the phase boundary between metal and AA insulator at small $U$ is qualitatively consistent with prior results obtained within the self-consistent Hartree approximation~\cite{Kerala2015}.

\begin{figure}[t]
    \centering
    \includegraphics[width=\columnwidth]{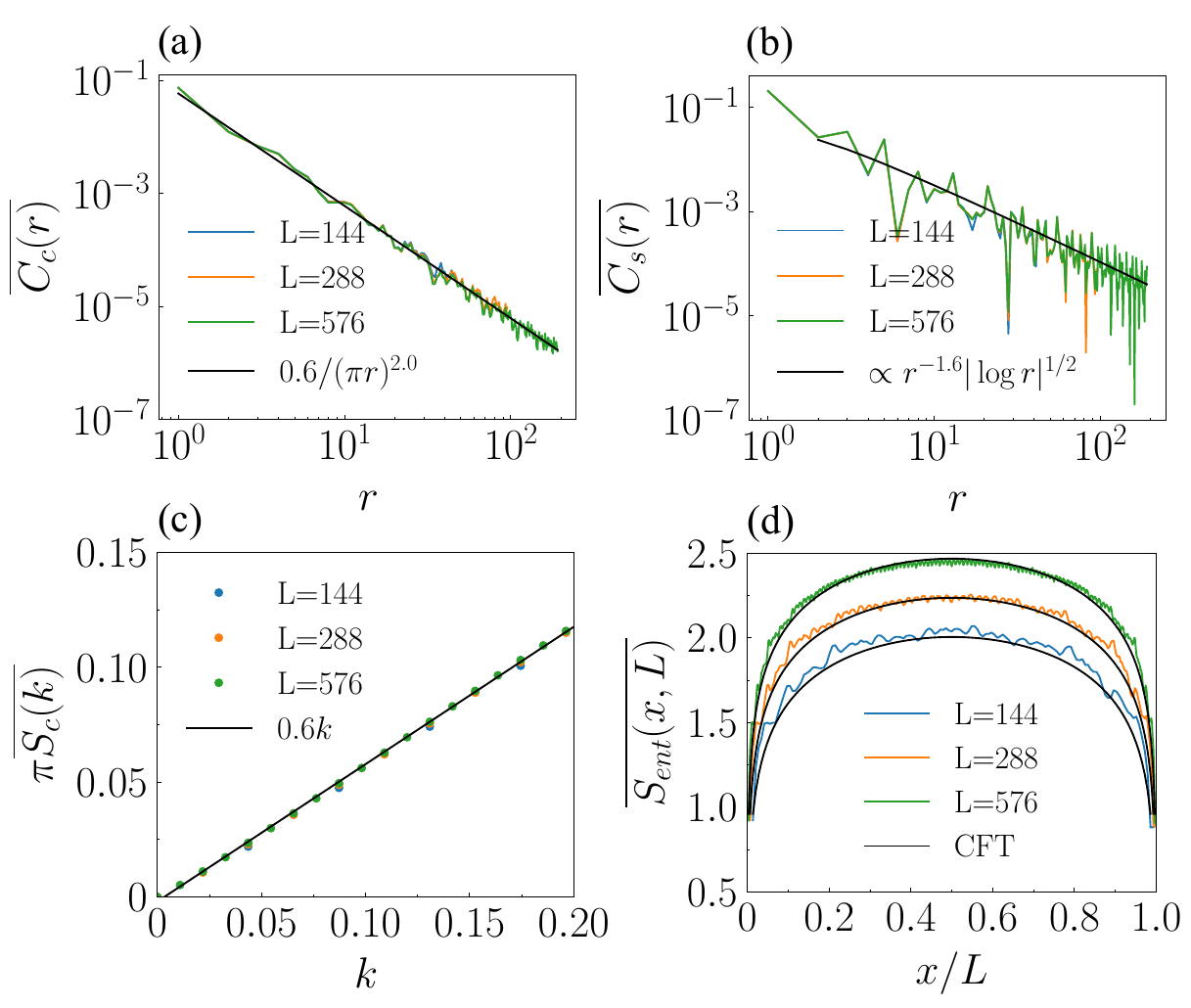}
    \caption{Luttinger liquid characteristics of the ground state at half filling, $\lambda/t = 4.4$ and $U/t = 6$. Data are shown for several system sizes $L$.
    (a)--(b) Phase-averaged and site-averaged charge correlation function $\overline{C_{\mathrm{c}}(r)}$ (panel a) and spin correlation function $\overline{C_{\mathrm{s}}(r)}$ (panel b) between sites $x$ and $x+r$ (averaged over $x$ from $L/3 - 5$ to $L/3+5$), with $0 \leq r \leq L/3$. Black lines are power-law fits $\propto r^{-2}$ and $\propto r^{-1.6} \, \lvert \log r \rvert^{1/2}$, respectively.
    (c) Phase-averaged charge structure factor $\overline{S_{\mathrm{c}}(k)}$ at small $k$, with a linear fit yielding the charge Luttinger parameter $K_{\mathrm{c}} = 0.6$.
    (d) Phase-averaged entanglement entropy profiles $\overline{S_{\text{ent}}(x,L)}$. Black curves are the Cardy formula~(\ref{eq:S_ent}) with $c = 2$ and additive constant $0.73$.
    \label{fig2}}
\end{figure}

Having established the existence of a metallic phase at intermediate values of $\lambda/t$ and $U/t$ in the AAH model~(\ref{eq:Ham}) at half filling, we now examine the physical characteristics of this phase in more detail. To this end, we select the parameters $\lambda/t=4.4$ and $U/t=6.0$ [the black triangle indicated by an arrow in Fig.~\ref{fig1}(e)], and study the charge and spin correlation functions,
\begin{align}
C_{\mathrm{c}}(x_j - x_l; x_l) &= 
\langle n_j n_l \rangle - \langle n_j \rangle \langle n_l \rangle , \\*
C_{\mathrm{s}}(x_j - x_l; x_l) &= 
\langle s^z_j s^z_l \rangle - \langle s^z_j \rangle \langle s^z_l \rangle ,
\end{align}
where $s^z_i \equiv n_{i\uparrow} - n_{i\downarrow}$. We choose reference points $x$ in a small window $L/3-5 \leq x \leq L/3+5$ and consider correlations between sites $x$ and $x+r$ for various relative coordinates $0 \leq r \leq L/3$, so that the data come from the middle of the chain. In Figs.~\ref{fig2}(a) and (b), we plot the phase- and site-averaged correlation functions $\overline{C_{\mathrm{c},\mathrm{s}}(r)} \equiv \frac{1}{11} \sum_{x = L/3-5}^{L/3+5}$ $\overline{C_{\mathrm{c},\mathrm{s}}(r;x)}$ versus $r$. The observed power laws, $\overline{C_{\mathrm{c}}(r)} \sim r^{-2}$ and $\overline{C_{\mathrm{s}}(r)} \sim r^{-1.6}$, are consistent with Luttinger liquid behavior, as we now discuss.

In a Luttinger liquid, all correlation functions are controlled by two Luttinger parameters, $K_{\mathrm{c}}$ (for charge) and $K_{\mathrm{s}}$ (for spin). As we have already mentioned, $K_{\mathrm{c}}$ can be extracted from the charge structure factor $S_{\mathrm{c}}(k) = L^{-1} \sum_{r,x} e^{ikr} C_{\mathrm{c}}(r;x)$ [Eq.~(\ref{eq:S_charge_k})] via
\begin{equation} 
K_{\mathrm{c}} = \lim_{k \rightarrow 0} \frac{\pi S_{\mathrm{c}}(k)}{k} .
\end{equation}
In Fig.~\ref{fig2}(c), we plot the phase-averaged structure factor $\pi \overline{S_{\mathrm{c}}(k)}$ versus $k$ (for small $k$); a linear fit yields $K_{\mathrm{c}} = 0.6$.

For a Luttinger liquid with $SU(2)$ spin rotation symmetry, $K_{\mathrm{s}} \equiv 1$ and the correlation functions have the form \cite{Giamarchi1989,Voit1988,Voit1995,Fradkin2013}
\begin{align}
\label{eq:C_charge}
C_{\mathrm{c}}(r) &= 
\frac{K_{\mathrm{c}}}{(\pi r)^2} + A_{\mathrm{c}} \frac{\cos(2 k_F r)}{\lvert r \rvert^{1+K_{\mathrm{c}}}} \, \lvert \log{\lvert r \rvert} \rvert^{-3/2} + \, \cdots , \\*
\label{eq:C_spin}
C_{\mathrm{s}}(r) &= 
\frac{1}{(\pi r)^2} + A_{\mathrm{s}} \frac{\cos(2 k_F r)}{\lvert r \rvert^{1+K_{\mathrm{c}}}} \, \lvert \log{\lvert r \rvert} \rvert^{1/2} + \, \cdots ,
\end{align}
where $k_F$ is the Fermi wavevector and $A_{\mathrm{c}}, A_{\mathrm{s}}$ are non-universal constants. Since $K_{\mathrm{c}} < 1$ (as is generally the case for repulsive interactions), the second term dictates the true asymptotic behavior as $\lvert r \rvert \to \infty$ in both Eqs.~(\ref{eq:C_charge}) and (\ref{eq:C_spin}). However, because of the different log corrections, there is a large intermediate regime where $C_{\mathrm{c}}(r) \sim 1/r^2$ is dominated by the first term in (\ref{eq:C_charge}) while $C_{\mathrm{s}}(r) \sim \lvert \log r \rvert^{1/2} / r^{1+K_{\mathrm{c}}}$ is dominated by the second term in (\ref{eq:C_spin}); this intermediate regime contains the range of system sizes accessible in our numerics. Using the value $K_{\mathrm{c}} = 0.6$ obtained from the charge structure factor, we reproduce the observed power laws: $\overline{C_{\mathrm{c}}(r)} \sim K_{\mathrm{c}} / (\pi r)^2$ [Fig.~\ref{fig2}(a)] and $\overline{C_{\mathrm{s}}(r)} \sim r^{-1.6} \, \lvert \log r \rvert^{1/2}$ [Fig.~\ref{fig2}(b)].

To further verify the Luttinger liquid character of the metallic phase, we compute the von Neumann entanglement entropy
\begin{equation} 
S_{\text{ent}}(x,L) = - \, \mathrm{tr}_x ( \rho_x \log \rho_x ) ,
\end{equation}
where $\rho_x$ denotes the reduced density matrix of the subsystem consisting of sites $i=1,2,\dots,x$. In Fig.~\ref{fig2}(d), we plot the phase-averaged entanglement entropy $\overline{S_{\text{ent}}(x,L)}$ as a function of $x/L$ for various system sizes $L$. The data are well fit by the 1+1D conformal field theory result \cite{Calabrese2009}
\begin{equation}
\label{eq:S_ent}
S_{\text{ent}}(x,L) = 
\frac{c}{6} \log\left( \frac{L}{\pi} \sin \frac{\pi x}{L}\right) + \mathrm{const.} ,
\end{equation}
with central charge $c = 2$, consistent with a Luttinger liquid with gapless charge and spin modes.

\begin{figure}[t]
    \centering
    \includegraphics[width=\columnwidth]{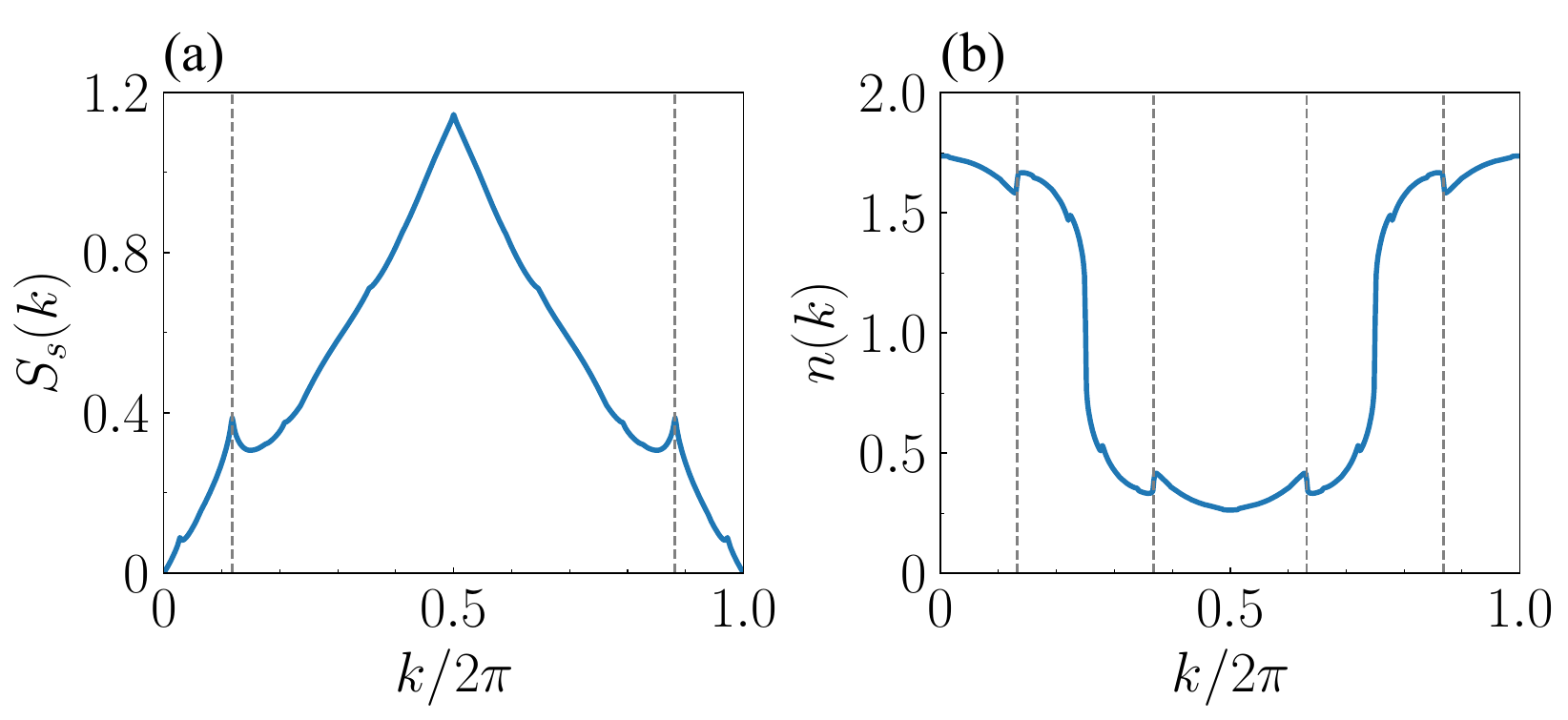}
    \caption{Signatures of the quasiperiodic potential in properties of the state at half filling, $\lambda/t = 4.4$ and $U/t = 6$.
    (a) The spin structure factor $S_{\mathrm{s}}(k)$. In addition to the usual peak at $k = \pi$, it displays smaller peaks at $k=\pi \pm q \mod 2\pi$ (marked by the gray dashed lines).
    (b) The electron distribution function $n(k)$. In addition to the usual power-law singularities at $k = \pm \pi/2$, it contains four smaller such features at $k = \pm \pi/2 \pm q \mod 2\pi$ (marked by gray dashed lines). Data is for $L=576$.
    \label{fig3}} 
\end{figure}

Although the metallic phase exhibits many characteristics of a Luttinger liquid, it does differ in some respects from a conventional clean Luttinger liquid. This distinction is evident in the spin structure factor,
\begin{equation} 
\label{eq:S_spin_k}
S_{\mathrm{s}}(k) = 
\frac{1}{L}\sum_{jl} e^{ik(x_j-x_l)} 
\left[ \langle s^z_j s^z_l \rangle - \langle s^z_j \rangle \langle s^z_l \rangle \right] ,
\end{equation}
which is plotted in Fig.~\ref{fig3}(a). In a conventional Luttinger liquid, $S_{\mathrm{s}}(k)$ features a single peak at $k=\pi$, corresponding to $2k_F$ of the half-filled chain. In contrast, the calculated $S_{\mathrm{s}}(k)$ of our quasiperiodic Luttinger liquid displays prominent additional peaks at $k = \pi \pm q$, where $q$ is the wavevector of the quasiperiodic potential [Eq.~(\ref{eq:V_i})]. 

The additional peaks in $S_{\mathrm{s}}(k)$ can be related to pseudo-Fermi points that are identifiable in the $k$-space electron distribution function,
\begin{equation} 
n(k) = \sum_{\sigma} \langle \hat{c}_{\sigma}^{\dagger}(k) \, \hat{c}_{\sigma}^{\phantom{\dagger}}(k) \rangle ,
\end{equation}
where $\hat{c}_{\sigma}(k) \equiv L^{-1/2} \sum_j e^{i k x_j} c_{j \sigma}$. The calculated distribution function is plotted in Fig.~\ref{fig3}(b). In addition to the usual power-law singularities at $k = \pm \pi/2$, $n(k)$ also contains four smaller such features at $k = \pm \pi/2 \pm q$. The additional peaks in $S_{\mathrm{s}}(k)$ can be crudely understood as arising from scattering between the latter pseudo-Fermi points and the former normal ones. A similar behavior is observed in the charge structure factor, $S_{\mathrm{c}}(k)$, in the non-interacting limit (see Appendix~\ref{app:S_charge}). However, at the present parameters, $S_{\mathrm{c}}(k)$ is a relatively smooth function of $k$ without sharp features, as shown in Fig.~\ref{figA3}(c).

\subsection*{Finite doping}

\begin{figure}[t]
    \centering
    \includegraphics[width=\columnwidth]{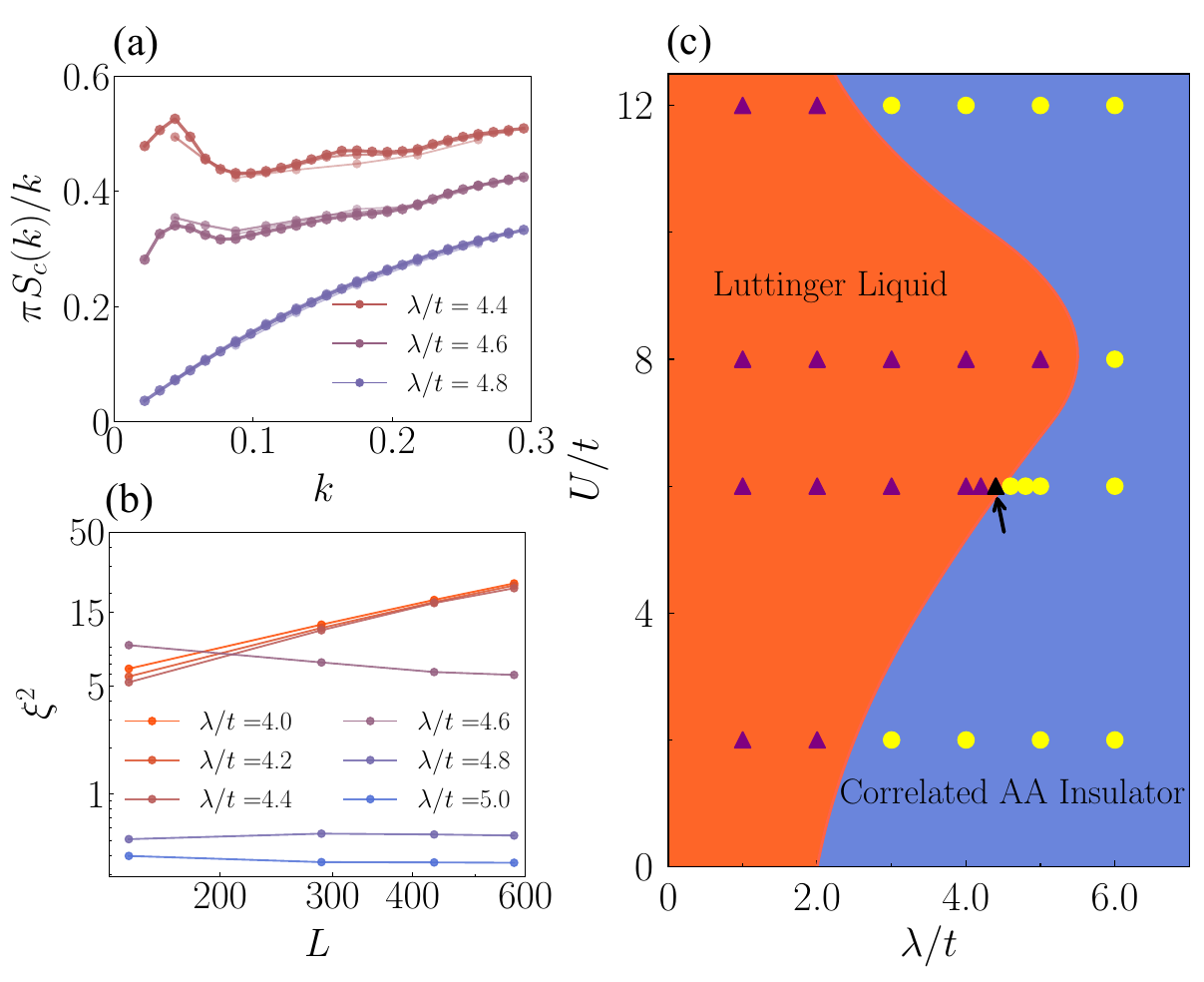}
    \caption{Results on the ground state phase diagram of the AAH model at $n = 11/12$ filling.
    (a) Behavior of the charge structure factor $S_{\mathrm{c}}(k)$ as $k \to 0$. 
    (b) Scaling of the charge localization length $\xi$ with system size $L$. Data are shown at $U/t = 6$ for various values of $\lambda/t$.
    (c) Estimated ground state phase diagram of the AAH model at $11/12$ filling. Symbols mark parameter values at which data were collected. These are identified as metallic (purple/black triangles) or insulating (yellow circles) according to the behavior of $S_{\mathrm{c}}(k)$ and $\xi$, as in the half-filled case. The precise shape of the phase boundaries is again conjectural, but is the simplest configuration consistent with our numerical results at the marked points and the known results in the $U=0$ or $\lambda=0$ limits.
    \label{fig4}} 
\end{figure}

Next, we study the effect of doping away from half-filling on the ground state phase diagram of the AAH model~(\ref{eq:Ham}). In the clean limit ($\lambda=0$, $U > 0$), any finite doping destroys the Mott insulator by moving the chemical potential above or below the Mott gap, and the ground state is a Luttinger liquid \cite{Lieb1968,Voit1995}. On the other hand, in the non-interacting limit ($U = 0$), there is still a transition from a metal at $\lambda/t < 2$ to an Aubry-Andr\'{e} insulator at $\lambda/t > 2$ \cite{Aubry1980}. We determine the boundary between these phases at nonzero $\lambda$ and $U > 0$ as in the half-filled case, by considering the long-wavelength behavior of the charge structure factor $S_{\mathrm{c}}(k)$ and the system-size dependence of the charge localization length $\xi$.

We study the model at $n = 11/12$ filling (i.e.~at $1/12$ hole doping). The results are summarized in Fig.~\ref{fig4}. Representative plots of $\pi S_{\mathrm{c}}(k)/k$ versus $k$, and of $\xi^2$ versus $L$, for $U/t = 6$ and various values of $\lambda/t$, are shown in Figs.~\ref{fig4}(a) and \ref{fig4}(b), respectively. Figure~\ref{fig4}(c) shows our obtained finite-doping phase diagram. At all values of $U > 0$, the system is a metal at small $\lambda/t$ and a correlated Aubry-Andr\'{e} insulator at sufficiently large $\lambda/t$. The critical $\lambda_{\mathrm{c}}/t$ at which the MIT occurs is a non-monotonic function of $U$, and is largest at $U/t \approx 8$ (where we find the metallic phase persists up to $\lambda/t \approx 5$). A similarly shaped phase boundary has been found separating ergodic and many-body localized phases in cold atom experiments \cite{Schreiber2015} and in the spinless AA model with nearest-neighbor repulsion \cite{Bera2017}.

The shape of the phase boundary can be understood intuitively on the basis of the half-filling phase diagram [Fig.~\ref{fig1}(e)]. Light hole doping will not fundamentally alter the state in the correlated AA insulating region, but it will introduce holes below the Mott gap throughout the quasiperiodic Mott insulating region. In the limit of large $U/t$, the holes are themselves weakly interacting quasiparticles and should therefore undergo an Aubry-Andr\'{e} localization transition at $\lambda_{\mathrm{c}}/t = 2$. These considerations lead to a phase diagram similar to Fig.~\ref{fig4}(c), with the extended metallic phase at intermediate $U/t$ corresponding to a delocalized-hole-doped region to the left of the metallic sliver in Fig.~\ref{fig1}(e).

\begin{figure}[t]
    \centering
    \includegraphics[width=\columnwidth]{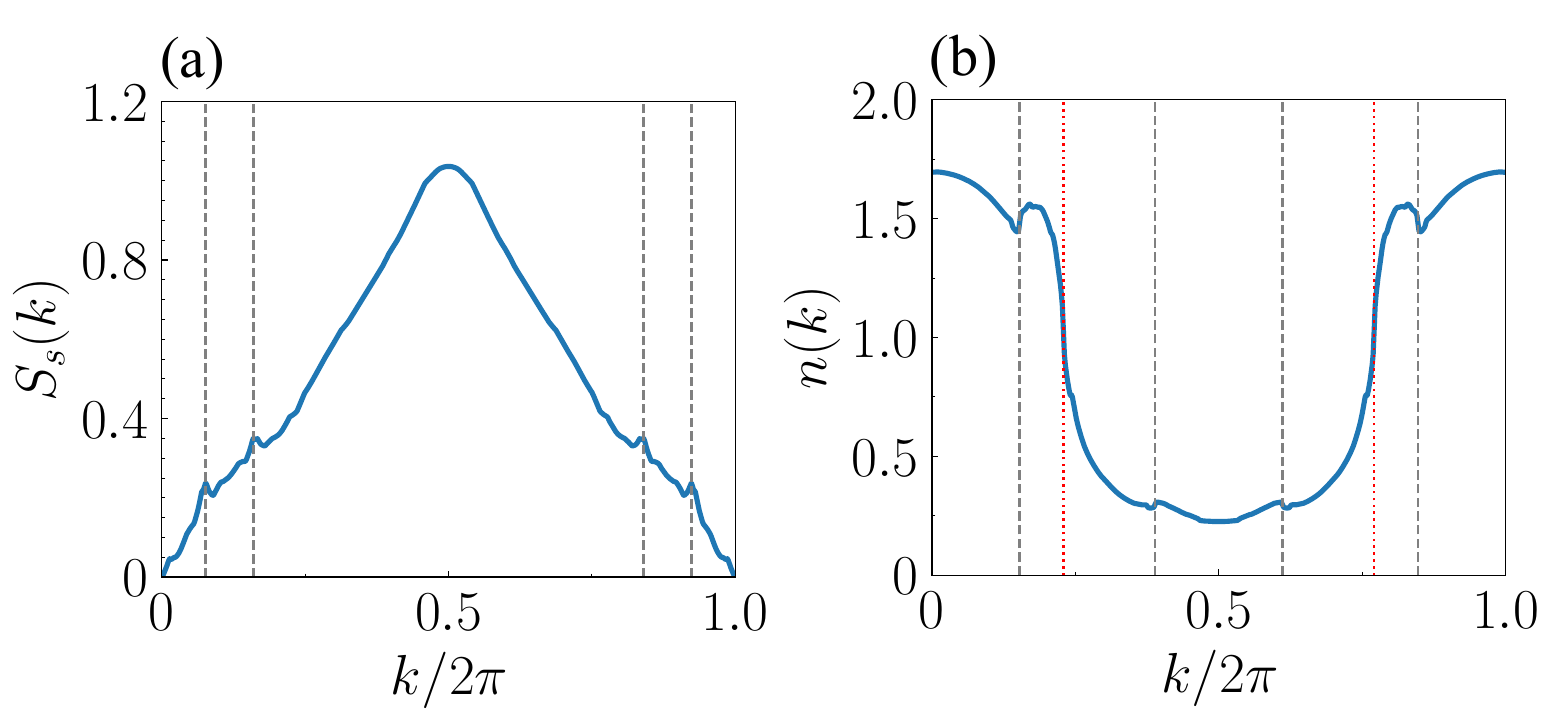}
    \caption{Signatures of the quasiperiodic potential in properties of the state at $n = 11/12$ filling, $\lambda/t = 4.4$ and $U/t = 6$.
    (a) The spin structure factor $S_{\mathrm{s}}(k)$. It displays peaks at $k = \pm n\pi \pm q \mod 2\pi$ (marked by the gray dashed lines).
    (b) The electron distribution function $n(k)$. In addition to the usual power-law singularities at $k = \pm n\pi/2 \mod 2\pi$ (marked by red dotted lines), it contains four smaller such features at $k = \pm n\pi/2 \pm q \mod 2\pi$ (marked by gray dashed lines). Data is for $L=576$.
    \label{fig5}} 
\end{figure}

As in the half-filled case, the lightly doped metallic phase exhibits signatures of the quasiperiodic potential in its spin structure factor $S_{\mathrm{s}}(k)$ and electron distribution function $n(k)$. Consider the doped system with $\lambda/t = 4.4$ and $U/t = 6.0$ [the black triangle indicated by an arrow in Fig.~\ref{fig4}(e)]. These are the same parameters for which we studied in detail the metal in the half-filled system. As shown in Fig.~\ref{fig5}(a), $S_{\mathrm{s}}(k)$ of the doped metal possesses peaks at $k=\pm n\pi \pm q$, in contrast to the peaks at $k=\pi \pm q$ in the half-filled metal [Fig.~\ref{fig3}(a)]. This peak splitting can be understood in terms of $n(k)$ of the doped system [Fig.~\ref{fig5}(b)], which contains Fermi points at $k = \pm n\pi/2$ and pseudo-Fermi points at $\pm n\pi/2 \pm q$ [cf.~Fig.~\ref{fig3}(b)]. These features clearly reflect the combined effects of doping and the quasiperiodic potential.

\section*{Conclusion}

We show that the interplay between a repulsive Hubbard interaction and quasiperiodic Aubry-Andr\'{e} potential in a 1D spinful fermion chain gives rise to an interaction-stabilized Luttinger liquid, both at and away from half filling. We characterize the Luttinger liquid through its charge and spin correlations, structure factors, and entanglement entropy. While this quasiperiodic metallic phase resembles the ``clean'' Luttinger liquid in many respects, including in its charge properties, the spin characteristics are altered due to extra scattering introduced by the quasiperiodic potential.

Our work thus establishes the existence of a metallic phase in a 1D model of electrons with strong interactions and a proxy for disorder (quasiperiodic potential), and sheds light on the necessary conditions---namely, sufficient competition between the two factors---to establish such a phase. Our results suggest that the 1D AAH model may also capture other interesting aspects of the interplay between disorder and correlations in higher dimensions; further theoretical and numerical studies exploring these aspects are warranted.

The AAH model considered in our work may also be studied experimentally, since quasi-random optical lattices are generated through incommensurate beam interference. The unusual metallic phase at half-filling can thus in principle be realized in a cold atom system \cite{Kondov2015,Schreiber2015}. The metallic phase should exhibit uniform Luttinger-liquid-like charge properties but quasiperiodically modulated magnetic correlations.

Finally, while our investigation was restricted to a particular form of quasiperiodic potential, we anticipate that our results should apply qualitatively to other forms as well. In particular, certain ``generalized Aubry-Andr\'{e} potentials'' admit a mobility edge \cite{Ganeshan2015}, allowing for the possibility of distinct metal--insulator transitions as a function of doping and interaction strength. The study of such generalized quasiperiodic potentials, as well as finite-temperature behavior and transport properties of the quasiperiodic metallic phase, are left for future work.


\begin{acknowledgements}

We gratefully acknowledge helpful discussions with Philip Crowley, Trithep Devakul, Chris Laumann, and especially Srinivas Raghu.
T.X.~and R.Z.C.~were supported by the National Natural Science Foundation of China under Grant No.~11888101. 
J.J.Y.~was supported by the National Science Foundation Graduate Research Fellowship under Grant No.~DGE-1656518.
C.M. was supported by the Gordon and Betty Moore Foundation's EPiQS Initiative through GBMF8686, and by the Department of Energy, Office of Basic Energy Sciences, under contract No.~DE-AC02-76SF00515.

\end{acknowledgements}

\bibliography{refs}

\clearpage
\appendix

\renewcommand{\thefigure}{A\arabic{figure}}
\renewcommand{\theequation}{A\arabic{equation}}
\setcounter{figure}{0}
\setcounter{equation}{0}

\section{Benchmark results for the non-interacting Aubry-Andr\'{e} model}
\label{app:AA_benchmark}

In this section, we study the non-interacting Aubry-Andr\'{e} model [the $U=0$ limit of (\ref{eq:Ham})] to benchmark our metrics which distinguish the metallic from insulating phases (cf.~the discussion in the Results section of the main text). We focus on quasiperiodic potential strengths $\lambda$ near the known MIT at $\lambda/t=2$.

\begin{figure}[t]
    \centering
    \includegraphics[width=\columnwidth]{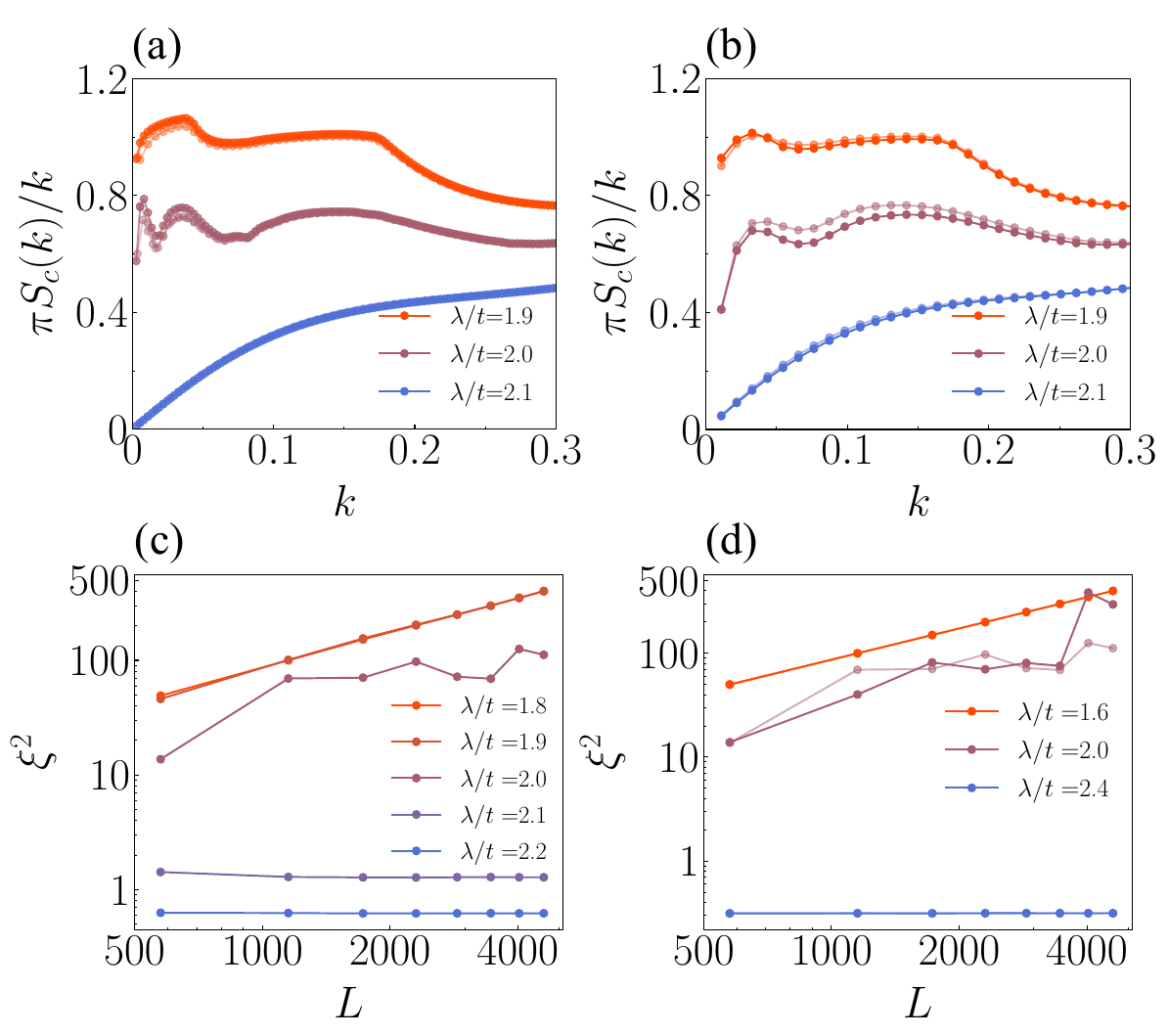}
    \caption{Results on the non-interacting Aubry-Andr\'{e} model for quasiperiodic potential strengths $\lambda$ near the known MIT at $\lambda/t = 2$.
    (a)--(b) Behavior of the charge structure factor $S_{\mathrm{c}}(k)$ as $k \to 0$. A nonzero extrapolated intercept of $\pi S_{\mathrm{c}}(k)/k$ indicates a metal, while a vanishing intercept indicates an insulator, as discussed in the main text. In panel a, the data are computed at a fixed phase $\phi = \pi/5$ of the quasiperiodic potential, and heavier lines correspond to increasing system sizes of $L = 1152$, $1728$, and $2304$ sites. In panel b, $L = 576$, and the lighter (heavier) lines represent data computed at $\phi = \pi/5$ (averaged over $10$ equally spaced phases, $\phi = 0, \pi/5, 2\pi/5, \dots, 9\pi/5$).
    (c)--(d) Scaling of the charge localization length $\xi$ with system size $L$. $\xi^2 \sim L^{\alpha}$ with $\alpha > 0$ in the metal, while $\xi^2 \sim L^0$ in the insulator. In panel c, the data are computed at $\phi = \pi/5$. In panel d, the lighter (heavier) lines represent data computed at $\phi = \pi/5$ (phase-averaged).
    \label{figA1}} 
\end{figure}

In Fig.~\ref{figA1}(a), we study the long-wavelength behavior of the charge structure factor $S_{\mathrm{c}}(k)$ by plotting $\pi S_{\mathrm{c}}(k)/k$ versus $k$ for $k > 0$. The extrapolated intercept yields $K_{\mathrm{c}} = 1$ in the metallic phase at $\lambda/t < 2$, and vanishes in the localized phase at $\lambda/t > 2$, as expected. At the transition, $\lambda/t = 2$, we also find a nonzero intercept $K_{\mathrm{c}} \approx 0.7$, which is consistent with previous results~\cite{Miguel2023}. We find that the results are only sensitive to the phase $\phi$ of the quasiperiodic potential very near the critical value $\lambda/t=2$, as shown in Fig.~\ref{figA1}(b).

In Fig.~\ref{figA1}(c), we study the scaling with system size $L$ of the charge localization length $\xi$ by plotting $\xi^2$ versus $L$. We find that $\xi^2 \sim L^{\alpha}$ with $\alpha = 1$ in the metallic phase at $\lambda/t < 2$, and $\alpha = 0$ in the localized phase at $\lambda/t > 2$, as expected (and consistent with previous results \cite{Kerala2015}). At the transition, $\lambda/t = 2$, $\xi^2$ still diverges with increasing $L$, but the dependence seems to be more complicated than a simple power law---perhaps because the single-particle spectrum and eigenstates are multifractal at the transition \cite{Wilkinson1984}. Again, the results are only sensitive to the phase $\phi$ of the quasiperiodic potential very near the critical point, as shown in Fig.~\ref{figA1}(d). Note that the charge localization length $\xi$ is not simply related to the localization lengths of single particle eigenstates in the non-interacting limit.

\renewcommand{\thefigure}{B\arabic{figure}}
\renewcommand{\theequation}{B\arabic{equation}}
\setcounter{figure}{0}
\setcounter{equation}{0}

\section{Distinguishing between Mott insulator and correlated Aubry-Andr\'{e} insulator}
\label{app:ins_transition}

In this section, we discuss how we distinguish between the Mott-like and Aubry-Andr\'{e}-like insulators at half-filling. Consider the energy difference between the ground state at exactly half-filling and the ground state with one electron added or removed. In the non-interacting Aubry-Andr\'{e} insulator ($U = 0$, $\lambda/t > 2$), the electron will be added to (removed from) a localized single-particle level just above (below) the Fermi level, so the energy difference should vanish in the thermodynamic limit. In contrast, in the pure Mott insulator ($U > 0$, $\lambda=0$), the electron must be added to (removed from) the upper (lower) Hubbard band, resulting in a finite energy difference set by the repulsion $U$. Note that, while the relative energy cost to add or remove an electron depends on the value of the chemical potential $\mu$ within the Mott gap, the average of the two energy costs does not depend on $\mu$ and directly measures the gap. Thus, we may define a single particle gap $\Delta_{\mathrm{p}} = (E_{N+1} - E_N)/2 + (E_{N-1} - E_N)/2$ and use this quantity to distinguish the two types of insulators; $\Delta_{\mathrm{p}}$ should be nonzero in the Mott insulator and zero in the AA insulator (in the thermodynamic limit). More precisely, since adding or removing an electron necessarily changes the total spin by a half-integer,
\begin{equation}
\Delta_{\mathrm{p}} \equiv 
\tfrac{1}{2} E_{N/2+1,N/2} + \tfrac{1}{2} E_{N/2-1,N/2} - E_{N/2,N/2} ,
\end{equation}
where $E_{N_{\uparrow}, N_{\downarrow}}$ denotes the energy of the ground state with $N_{\uparrow}$ spin-up electrons and $N_{\downarrow}$ spin-down electrons. We may similarly define the charge gap,
\begin{equation}
\label{eq:charge_gap}
\Delta_{\mathrm{c}} \equiv 
\tfrac{1}{2} E_{N/2+1,N/2+1} + \tfrac{1}{2} E_{N/2-1,N/2-1} - E_{N/2,N/2} ,
\end{equation}
the average energy cost of adding or removing two electrons with net spin zero, and use this quantity to distinguish the insulators. For technical reasons, we find it preferable to use $\Delta_{\mathrm{c}}$ rather than $\Delta_{\mathrm{p}}$.

\begin{figure}[t]
    \centering
    \includegraphics[width=\columnwidth]{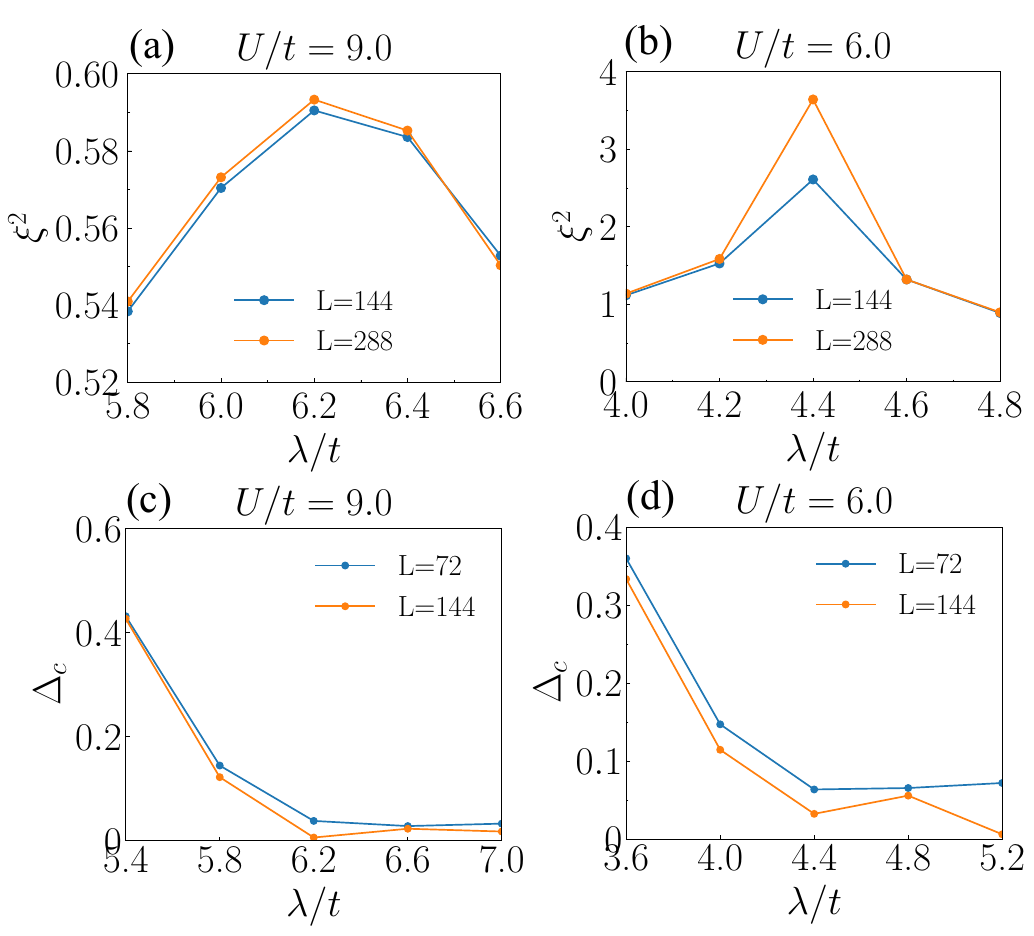}
    \caption{Data showing the transition from the gapped Mott insulator to the gapless correlated Aubry-Andr\'{e} insulator upon varying $\lambda/t$ at large $U/t > 0$.
    (a)--(b) The charge localization length $\xi$ as a function of $\lambda/t$ at $U/t = 9$ (panel a) and $U/t=6$ (panel b), for two system sizes $L$. $\xi^2$ increases with $L$ (metallic behavior) near $\lambda/t \approx 4.4$ at $U/t = 6$, but is $L$-independent (insulating) at all other parameters shown.
    (c)--(d) The charge gap $\Delta_{\mathrm{c}}$ [Eq.~(\ref{eq:charge_gap})] as a function of $\lambda/t$ at $U/t = 9$ (panel c) and $U/t=6$ (panel d). The system appears to be gapless (i.e.~$\Delta_{\mathrm{c}} \to 0$ with increasing $L$) for $\lambda/t \gtrsim 6.2$ at $U/t = 9$, and for $\lambda/t \gtrsim 4.4$ at $U/t = 6$, but appears to be gapped ($\Delta_{\mathrm{c}} > 0$) at smaller values of $\lambda/t$. In order to alleviate finite size effects, we calculate $\Delta_{\mathrm{c}}$ for 30 choices of the phase $\phi$ of the quasiperiodic potential and report the minimum, for each set of parameters.
    \label{figA2}} 
\end{figure}

Figure~\ref{figA2} shows how the charge localization length $\xi$ and charge gap $\Delta_{\mathrm{c}}$ behave as the half-filled chain evolves between the gapped Mott insulator and gapless Aubry-Andr\'{e} insulator with varying $\lambda/t$ at fixed $U/t$. Results for $U/t = 9$ are shown in Figs.~\ref{figA2}(a) and (c). Here, there is no intervening metallic phase, as evidenced by $\xi$ converging to a finite value with increasing system size for all values of $\lambda/t$. Meanwhile, the charge gap $\Delta_{\mathrm{c}}$ is nonzero for $\lambda/t \lesssim 6$ but vanishes with increasing $L$ for $\lambda/t \gtrsim 6$, signalling a direct transition from the Mott insulator to the correlated Aubry-Andr\'{e} insulator. Note that $\xi$ varies non-monotonically with $\lambda/t$, and is largest (although still small) near the insulator-insulator transition at $\lambda/t \approx 6.2$. Similar data are shown for $U/t=6$ in Figs.~\ref{figA2}(b) and (d). In this case, there is an intervening metallic phase around $\lambda/t = 4.4$, as indicated by $\xi$ increasing with $L$ (see also the extensive results in the main text). $\Delta_c$ is nonzero in the Mott insulator at smaller $\lambda/t$, but vanishes with increasing $L$ both in the metal and in the correlated Aubry-Andr\'{e} insulator at larger $\lambda/t$.

\renewcommand{\thefigure}{C\arabic{figure}}
\renewcommand{\theequation}{C\arabic{equation}}
\setcounter{figure}{0}
\setcounter{equation}{0}

\section{Charge structure factor}
\label{app:S_charge}

\begin{figure}[t]
    \includegraphics[width=\columnwidth]{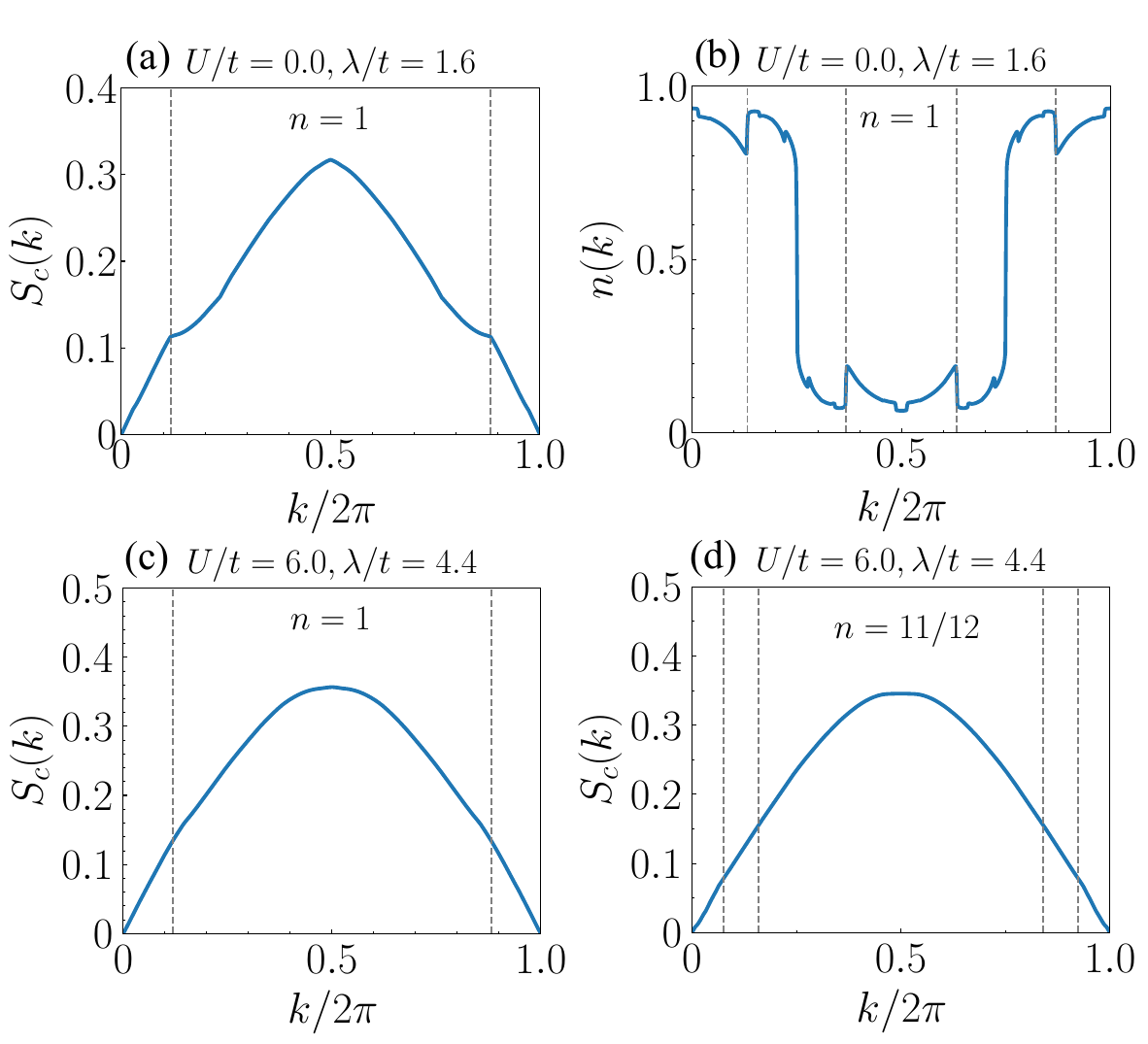}
    \caption{Signatures of the quasiperiodic potential in the charge structure factor $S_{\mathrm{c}}(k)$, or the lack thereof.
    (a) $S_{\mathrm{c}}(k)$ for the non-interacting half-filled system in the metallic phase. In addition to the usual peak at $k = \pi$, it displays kinks at $k=\pi \pm q \mod 2\pi$ (marked by the gray dashed lines).
    (b) The corresponding electron distribution function $n(k)$. In addition to the usual power-law singularities at $k = \pm \pi/2$, it contains four smaller such features at $k = \pm \pi/2 \pm q \mod 2\pi$ (marked by gray dashed lines).
    (c)--(d) $S_{\mathrm{c}}(k)$ for the interacting metal at half filling (panel c) or at $n=11/12$ filling (panel d). In both cases, $S_{\mathrm{c}}(k)$ is a relatively smooth function of $k$. The gray dashed lines mark $k=\pi \pm q \mod 2\pi$ (panel c) or $k = \pm n\pi \pm q \mod 2\pi$ (panel d). Data is for $L = 576$.
    \label{figA3}} 
\end{figure}

In this section, we discuss the charge structure factor $S_{\mathrm{c}}(k)$ in the metallic phase. In the non-interacting half-filled system, $S_{\mathrm{c}}(k)$ features two kinks at $k = \pi \pm q$, in addition to the usual peak at $k = \pi$, as shown in Fig.~\ref{figA3}(a). Much like the additional peaks at $k = \pi \pm q$ in the spin structure factor $S_{\mathrm{s}}(k)$ discussed in the main text, these kinks reflect the quasiperiodic potential, and can be crudely understood as arising from scattering between normal and pseudo-Fermi points which are identifiable in the $k$-space electron distribution function $n(k)$ [Fig.~\ref{figA3}(b)] at $k = \pm \pi/2$ and $k=\pm \pi/2 \pm q$, respectively. However, the features in $S_{\mathrm{c}}(k)$ are significantly smoothed out by the Hubbard interaction, and are almost invisible when $U/t = 6$, both at half-filling [Fig.~\ref{figA3}(c)] and at finite doping [Fig.~\ref{figA3}(d)]. This is in contrast to the features in $S_{\mathrm{s}}(k)$, which remain prominent in the interacting system [cf.~Fig.~\ref{fig3}(a) and Fig.~\ref{fig5}(a)].

\clearpage

\end{document}